\documentclass[dvips]{article}

\usepackage{icrc2011}

\title{Observation of the Perseus galaxy cluster with the MAGIC telescopes}
\newcommand{\etal}{\MakeLowercase{\textit{et al. }}} 
\shorttitle{Lombardi \etal Observation of the Perseus cluster with MAGIC}

\authors{
         Saverio Lombardi$^{1}$, 
         Fabio Zandanel$^{2}$, 
         Pierre Colin$^{3}$, 
         Michele Doro$^{4}$, 
         Dorothee Hildebrand$^{5}$, 
         Francisco Prada$^{2}$ 
         for the MAGIC Collaboration, 
         and Christoph Pfrommer$^{6}$, 
         Anders Pinzke$^{7}$
         }
\afiliations{
             $^1$Universit\`{a} di Padova and INFN, I-35131 Padova, Italy\\ 
             $^2$Instituto de Astrof\'{i}sica de Andaluc\'{i}a (IAA-CSIC), E-18080 Granada, Spain\\
             $^3$Max-Planck-Institut f\"{u}r Physik, D-80805 M\"{u}nchen, Germany\\
             $^4$Universitat Aut\`{o}noma de Barcelona, E-08193 Bellaterra, Spain\\
             $^5$ETH Zurich, 8093 Zurich, Switzerland\\
             $^6$Heidelberg Institute for Theoretical Studies (HITS), D-69118 Heidelberg, Germany\\
             $^7$UCSB, California, USA\\
             }
\email{saverio.lombardi@pd.infn.it}

\abstract{
The MAGIC ground-based Imaging Cherenkov experiment observed the Perseus galaxy cluster for a 
total of about 25~hr between November and December 2008 in single telescope mode and for nearly 
90~hr between October 2009 and February 2011 in stereoscopic mode. This survey represents the 
deepest observation of a cluster of galaxies at very high energies ever. It resulted in the 
detection of the central radio galaxy NGC~1275 and the head-tail galaxy IC~310. It also permits 
for the first time to put constraints on emission models predicting $\gamma$-rays from cosmic 
ray acceleration in the cluster and to investigate dark matter scenarios. Here, we will report 
the latest MAGIC results on these studies.
}

\keywords{Cluster of galaxies, Perseus, very high energy $\gamma$-rays, Cosmic rays, Dark Matter, MAGIC}

\begin{document}
\maketitle

\section{Introduction}
In the cosmological hierarchic clustering model \cite{peebles1993}, large-scale structures grow 
hierarchically through merging and accretion of smaller systems into larger ones, and thus clusters 
of galaxies are the latest objects to form. These astrophysical environments represent therefore 
the largest and most massive gravitationally bound systems in the Universe, with radii of few Mpc 
and total masses $M \sim (10^{14}-10^{15})M_{\odot}$, of which galaxies, gas, and dark matter (DM) 
contribute roughly for 5$\%$, 15$\%$ and 80$\%$, respectively (see e.g. \cite{sarazin1988,voit2005} 
for general overviews).\\
Cluster of galaxies are expected to be significant $\gamma$-ray emitters on the following general 
grounds. 
(i) Clusters are actively evolving objects and being assembled today, in the latest and most energetic 
phase of hierarchical structure formation \cite{forman2003}, they should dissipate energies of the 
order of the final gas binding energy through merger and accretion shocks as well as turbulences, 
which are also likely to accelerate non-thermal electrons and protons to high energies \cite{sarazin2002,minati2003,pfrommer2008}.
(ii) Clusters are home of different types of energetic outflows of powerful sources such as radio 
galaxies \cite{forman2003} and supernova-driven galactic winds \cite{volk2003}, and the intra-cluster 
medium (ICM) can function as an efficient energy reservoir.
(iii) Clusters contain large amounts of gas with embedded magnetic fields -- the ICM is known to be permeated 
by magnetic fields with strengths $B \sim 1-10$~$\mu$G \cite{vogt2005} -- often showing direct evidence 
for shocks and turbulence as well as relativistic particles \cite{feretti2003}.\\
Additionally, galaxy clusters present very large mass--to--light ratios and considerable DM overdensities, 
which are considered crucial for indirect DM searches \cite{masc2011,pinzke2011}. Despite the fact that they are 
not as near as other potential DM candidates, like for instance the dwarf spheroidal galaxies 
\cite{albert2008,aliu2009,aleksic2011}, the large DM masses of clusters could make them ideal laboratories 
for the search of emissions in the $\gamma$-ray regime from DM annihilation \cite{masc2011,pinzke2011} 
or decay \cite{cuesta2011}. However, the recently underlined very extended nature of the DM signal in 
clusters \cite{masc2011,pinzke2011} is a major issue for the current generation of Cherenkov telescopes.\\
%
%
\section{The Perseus cluster and its observation with the MAGIC telescopes}
The Perseus cluster, at a distance of 77.7~Mpc (z~=~0.018), is the brightest X-ray cluster \cite{edge1992}, 
hosting a massive cooling flow and a luminous radio mini-halo that fills a large fraction of the cluster 
core region \cite{pedlar1990}. The radio mini-halo is well explained by the hadronic scenario where the 
radio emitting electrons are produced in hadronic CR proton-proton interactions with ICM protons requiring 
only a very modest fraction of a few percent of CR-to-thermal pressure \cite{pfrommer2004}.\\
The Perseus galaxy cluster was carefully chosen over other nearby clusters as it is the most promising 
target for the detection of $\gamma$-rays coming from the neutral pion decays result of the hadronic cosmic 
rays (CR) interactions with the ICM \cite{pinzke2010}. Moreover, the huge DM content of the cluster represents 
an additional strong motivation. Additionally, the central radio galaxy NGC~1275 is a very interesting GeV-TeV 
target, and hence it is a further reason for the observation of this cluster \cite{aleksic2010a} at very high 
energies (VHE). \\
The MAGIC (Major Atmospheric Gamma-ray Imaging Cherenkov) experiment consists of two 17~meter Imaging Air Cherenkov
Telescopes (IACTs) located on the Canary Island of La Palma (28.8$^{\circ}$N,~17.8$^{\circ}$W), 2200~meters above 
the sea level. The MAGIC telescopes are currently the largest world-wide existing IACTs. Since fall 2009 the 
telescopes are working together in stereoscopic mode which ensures an excellent sensitivity of $\sim$0.8$\%$ of Crab 
Nebula flux above $\sim$250~GeV in 50~hr of observations and an analysis threshold of $\sim$50~GeV \cite{carmona2011}.\\
The MAGIC experiment conducted the deepest survey ever made at VHE of the Perseus cluster, collecting data 
both in single telescope mode ($\sim$25~hr of MAGIC-I observations, between November and December 2008 
\cite{aleksic2010a}) and in stereoscopic mode ($\sim$90~hr of observations, between October 2009 and 
February 2011). The data were taken during dark nights at zenith angles below 35$^{\circ}$ (which guarantees 
a low analysis energy threshold) in false-source tracking (wobble) mode \cite{fomin1994}, in which the pointing 
direction alternates every 20 minutes between different positions symmetrically offset by 0.4$^{\circ}$ from 
the source. The data analysis was performed using the standard MAGIC analysis and reconstruction software 
\cite{lombardi2011}. These observations resulted in the detection of VHE emission from the 
head-tail radio galaxy IC~310 \cite{aleksic2010b} as well as the central radio galaxy NGC~1275 (ATel$\#2916$). 
A discussion of these two discoveries is presented elsewhere in this conference \cite{NGC1275_IC310_proc}. 
%
%
\section{Results and Discussion}
\subsection{MAGIC-I observations}
\begin{figure}[hbt!]
\vspace{5mm}
\centering
\includegraphics[width=3.in]{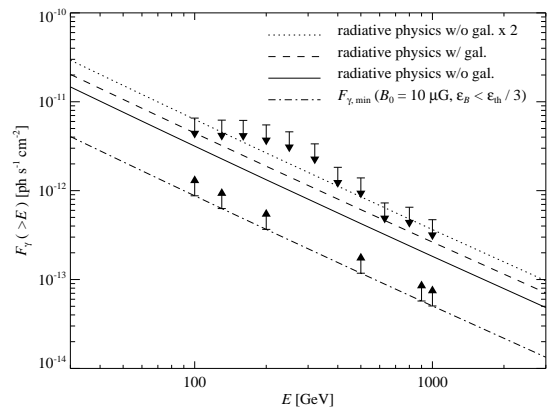}
\caption{
Comparison between the MAGIC-I observation integral flux upper limits (upper arrows) and the simulated integrated 
spectra of the CR induced $\gamma$-ray emission of the Perseus cluster \cite{pinzke2010}. The conservative model 
without galaxies (solid) is contrasted to the model with galaxies (dashed) and it is scaled with a factor of 2 
so that it is just consistent with the upper limits obtained in \cite{aleksic2010a} (dotted). Additionally shown 
is the minimum $\gamma$-ray flux estimated for the hadronic model of the radio mini-halo of the Perseus cluster 
(dash-dotted with arrows). Note that a non-detection of $\gamma$-rays at this level seriously challenges the 
hadronic model. See \cite{aleksic2010a} for details.
}
\label{fig1}
\end{figure}

MAGIC-I observations of the Perseus cluster took place during November and December 2008 resulting in a total 
effective time, after data selection, of 24.4~hr \cite{aleksic2010a}. No significant excess was found in the data.\\
The integral flux upper limits computed for different energy thresholds and for a supposed power law spectrum with 
spectral index of -2.2, as expected for the CR induced $\gamma$-ray emission in the energies of interest here 
\cite{pinzke2010}, are reported in Table \ref{tab1}.
\begin{table}[hbt!]
\begin{center}
\begin{tabular}{cc}
E$_{\mathrm{th}}$ [GeV] & $\Phi^{\mathrm{UL}}$ [$\times 10^{-12}$~cm$^{-2}$~s$^{-1}$]\\
\hline
100 &  6.55 \\
130 &  6.21 \\
160 &  6.17 \\
200 &  5.49 \\
250 &  4.59 \\
320 &  3.36 \\
400 &  1.83 \\
500 &  1.39 \\
630 &  0.72 \\
800 &  0.65 \\
1000 & 0.47 \\
\hline
\end{tabular}
\caption{
MAGIC-I observation integral flux upper limits, at $95\%$ confidence level, for a power-law $\gamma$-ray spectrum 
with spectral index of -2.2 above different energy threshold E$_{\mathrm{th}}$.
}
\label{tab1}
\end{center}
\end{table}
These upper limits are compared to the simulated flux of the $\gamma$-ray emission from decaying neutral pions 
that result from hadronic CR interactions with the ICM \cite{pinzke2010} in Figure~\ref{fig1}. The upper limits 
are a factor of two above our conservative model implying consistency with the cluster cosmological simulations 
of \cite{pinzke2010}, allowing to constrain the average CR-to-thermal pressure to $<4\%$ for the cluster core 
region and to $<8\%$ for the entire cluster. Using simplified assumptions adopted in earlier work, i.e.~a 
power-law spectrum with an index of $-2.1$, and a constant CR-to-thermal pressure for the peripheral cluster 
regions while accounting for the adiabatic contraction during the cooling flow formation, we would limit the 
ratio of cosmic ray-to-thermal energy to E$_{\mathrm{CR}}$/E$_{\mathrm{th}}<3\%$.
\\
The MAGIC-I data were also interpreted in terms of a potential DM annihilation emission \cite{aleksic2010a}. 
The Perseus cluster DM density profile was modeled with a typical Navarro-Frank-White (NFW) profile \cite{nfw1997} 
and the DM particle was assumed to be the \emph{neutralino} within the mSUGRA scenario \cite{chamseddine1982}, 
considering one of the most optimistic value for the particle physics factor for the energies of interest here 
\cite{masc2007}. The resulting expected $\gamma$-ray flux above the energy threshold of 100~GeV is then 
$\Phi_{\mathrm{DM, th}} =1.4 \times10^{-16}$~cm$^{-2}$~s$^{-1}$. The corresponding integral flux upper limit, 
computed for a generic DM annihilation spectrum modeled as a power-law with spectral index of -1.5, is 
$\Phi^{\mathrm{UL}}$($>$100~GeV)$=4.63 \times 10^{-12}$~cm$^{-2}$~s$^{-1}$. This means that a boost in the flux 
of the order of $10^4$ is needed to reach the predicted DM annihilation flux. This boost factor could come from 
different mechanisms (not taken into account in the calculation of $\Phi_{\mathrm{DM, th}}$) that may enhance 
the annihilation $\gamma$-ray flux notably, such as the presence of DM substructures \cite{kuhlen2008,springel2008} 
and the Sommerfeld effect \cite{lattanzi2009,pinzke2009}.\\ 
However, the above results, obtained without considering the contribution of DM substructures, should be substantially 
revised after two recent works \cite{masc2011,pinzke2011} that highlighted the strong contribution of subhaloes, 
which make the DM density profile extremely flat up to large distances from the core for all clusters 
(1.2$^{\circ}$ in the case of Perseus).
%
%
\subsection{Stereo observations}
\begin{figure}[hbt!]
\vspace{5mm}
\centering
\includegraphics[width=3.in]{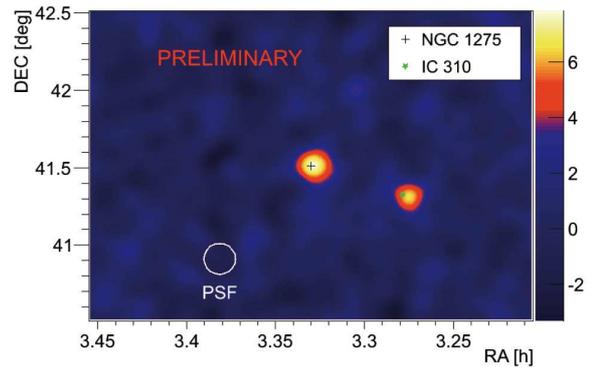}
\caption{
MAGIC stereo significance skymap of the Perseus cluster region above 150~GeV. For this map the overall 
stereo data sample of about $90$~hr have been used. NGC~1275 is clearly detected at the center of the 
cluster (ATel$\#2916$) and IC~310 is also visible in the map \cite{aleksic2010b}.
}
\label{fig2}
\end{figure}

\begin{figure}[hbt!]
\vspace{5mm}
\centering
\includegraphics[width=3.in]{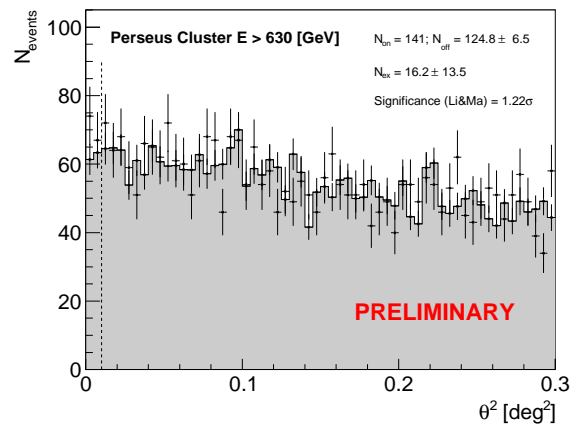}
\caption{
MAGIC stereo $\theta^2$ plot above 630~GeV. The black points represent the \emph{signal} while the gray 
shaded region is the \emph{background}. The vertical dotted line represents the \emph{signal region} where 
a point-like emission is expected. No significant $\gamma$-ray excesses are seen above that energy.
}
\label{fig3}
\end{figure}

The MAGIC telescopes observed the Perseus galaxy cluster in stereoscopic mode between October 2009 and 
February 2011 for a total effective time, after data selection, of about 90~hr. The observation resulted 
in the detection of VHE emission from the head-tail radio galaxy IC~310 \cite{aleksic2010b} as well as 
the central radio galaxy NGC~1275 (ATel$\#2916$) which are shown in the significance skymap above 150~GeV of Figure~\ref{fig2} 
(IC~310 and NGC~1275 discoveries are presented elsewhere in this conference \cite{NGC1275_IC310_proc}).
\\
To investigate a possible signal from CR hadronic interactions, we limit the analysis to energies where 
the central radio galaxy NGC~1275 is not emitting, i.e. approximately above 600~GeV. This is shown in 
the $\theta^2$ plot calculated from the overall stereo data sample in Figure~\ref{fig3}. We conclude that 
no CR induced emission is detected at energies above 630~GeV, where we only observe background fluctuations. 
The preliminary integral flux upper limit above that energy and assuming a power law spectrum with spectral index of $-2.2$
is $\Phi^{\mathrm{UL}} = 0.3 \times 10^{-12}$~cm$^{-2}$~s$^{-1}$.
\\
With the nearly 90~hr of stereo observation, we are now able to significantly tighten the constraints 
on the CR-to-thermal pressure obtained with the MAGIC-I observation \cite{aleksic2010a}. For the first 
time, we can put strong constraints on the hadronic model of radio mini-halo emission, potentially
starting to probe the acceleration physics of CRs at structure formation shocks. In fact, using the 
preliminary integral flux upper limit above 630~GeV, we would be able to constrain a combination of 
the assumed acceleration efficiency and CR transport parameters in the conservative model 
\emph{radiative physics without galaxies}, shown with a solid line in Figure~\ref{fig1}. However, 
we caution that this upper limit is only a preliminary estimate. A publication is in preparation and will be available soon.\\
%
%
\section{Summary}
We present the results achieved so far from the deep survey of the Perseus cluster of galaxies at VHE 
carried out by the MAGIC experiment, both in mono ($\sim$25~hr) and stereo ($\sim$90~hr) data taking mode.\\
Mono observations permitted to constrain the average CR-to-thermal pressure to $<4\%$ for the cluster 
core region and to $<8\%$ for the entire cluster. Using simplified assumptions adopted in earlier work 
(i.e.~a power-law spectrum with index of $-2.1$ and constant CR-to-thermal pressure for the peripheral 
cluster regions while accounting for the adiabatic contraction during the cooling flow formation) would 
allow us to limit the ratio of cosmic ray-to-thermal energy to E$_{\mathrm{CR}}$/E$_{\mathrm{th}}<3\%$.\\
Stereo observations permit to significantly tighten the previous constraints. This is made possible 
due to the fact that detection of the central cluster radio galaxy NGC~1275 \cite{NGC1275_IC310_proc} does 
not affect the higher energies (in fact, no signal is detected above approximately 600~GeV). This enables 
us to start to probe the acceleration physics of CRs at structure formation shocks. The estimation and 
interpretation of the flux upper limits, however, are still ongoing and a corresponding paper is under preparation.\\
%
%
\section{Acknowledgements}
We would like to thank the Instituto de Astrof\'{\i}sica de Canarias for the excellent working conditions 
at the Observatorio del Roque de los Muchachos in La Palma. The support of the German BMBF and MPG, 
the Italian INFN, the Swiss National Fund SNF, and the Spanish MICINN is gratefully acknowledged. 
This work was also supported by the Marie Curie program, by the CPAN CSD2007-00042 and MultiDark 
CSD2009-00064 projects of the Spanish Consolider-Ingenio 2010 programme, by grant DO02-353 of the 
Bulgarian NSF, by grant 127740 of the Academy of Finland, by the YIP of the Helmholtz Gemeinschaft,
by the DFG Cluster of Excellence ``Origin and Structure of the Universe'', by the DFG Collaborative 
Research Centers SFB823/C4 and SFB876/C3, and by the PolishMNiSzWgrant 745/N-HESSMAGIC/2010/0.
%
%



\clearpage

\end{document}